\documentclass[reprint,
groupedaddress,
showpacs,preprintnumbers,
 amsmath,amssymb,
 aps,
prl
]{revtex4-1}
\usepackage{graphicx}% Include figure files
\usepackage{dcolumn}% Align table columns on decimal point
\usepackage{bm}% bold math
\usepackage{mathrsfs}

\begin{document}

\title{Anomalous Josephson effect in $d$-wave superconductor junctions on TI surface}
\author{Bo Lu$^{1}$, Keiji Yada$^{1}$, A. A. Golubov$^{2,3}$, Yukio Tanaka$^{1,3}$}

\affiliation{$^1$~Department of Applied Physics, Nagoya University, Nagoya 464-8603, Japan\\
$^2$~Faculty of Science and Technology and MESA+ Institute for Nanotechnology,
University of Twente, 7500 AE Enschede, The Netherlands\\
$^3$~Moscow Institute of Physics and Technology, Dolgoprudny, Moscow 141700, Russia
}
\date{\today}

\begin{abstract}
We study Josephson effect of $d$-wave superconductor (DS)/ferromagnet
insulator(FI)/DS junctions on a surface of topological insulator (TI).
We calculate Josephson current $I\left(\varphi \right) $
for various orientations of the
junctions where $\varphi $ is the macroscopic phase difference between two
DSs. In certain configurations, we find anomalous current-phase relation $I(\varphi)=-I\left( -\varphi +\pi
\right)$ with $2\pi$ periodicity.
In the case where the
first order Josephson coupling is absent without magnetization in FI, $I(\varphi)$ can be proportional to
$\cos \varphi$.
The magnitude of the obtained Josephson current is enhanced
due to the zero energy states on
the edge of DS on TI. Even if we introduce an $s$-wave component of pair
potential in DS, we can still expect the anomalous current-phase relation in
asymmetric
DS junctions with $I\left( \varphi =0\right) \neq 0$. This can be used to
probe the induced $d$-wave component of pair potential
on TI surface in high-$T_{c}$ cuperate/TI hybrid structures.
\end{abstract}

\pacs{74.45.+c, 74.50.+r, 74.20.Rp}
\maketitle

%Title of the article

\section{I, Introduction}

Josephson effect has been a fundamental and central topic in
superconductivity and  contributed to determine the
pairing symmetry in unconventional superconductors
\cite{Tsuei,Tanaka2000,GKI}. It is well known that
standard current-phase relation (CPR) of Josephson current $I(\varphi)$ between
two superconductors
is $I(\varphi) \sim \sin \varphi $, where $\varphi $ is the macroscopic
phase difference. For $d$-wave superconductor (DS)
junctions, due to the presence of Andreev bound state at the interface \cite{CR,Kashiwaya95,Tanaka95},
exotic quantum interference effects exist.
One is the  non-monotonic temperature dependence of maximum Josephson current
\cite{Tanaka96,Barash,TanakaD97,Ilichev,Testa}
and second is the anomalous CPR \cite{Tanaka96,TanakaD97,Ilichev,Testa}.
Due to the presence of ABS, $\sin 2\varphi$ component of $I(\varphi)$
is enhanced and free energy of the junction can locate neither
$\varphi=0$ nor $\pm \pi$. Furthermore,
a pure $\sin 2\varphi $ CPR is possible for
$d_{x^{2}-y^{2}}$-wave /$d_{xy}$-wave superconductor junction\cite{Tanaka96,TanakaD97}.
Thus, $d$-wave junctions have really rich current phase relation
and its functionalities worth for
further research.

On the other hand, 3D topological insulators (TIs) \cite%
{Fu07prb,ZSC09,Hsieh08,Hsieh09,Hasan09,Chen09,Xia09,Ando10} is a material
with a topologically protected surface state due to the strong spin-orbit coupling. The generation of superconductivity on the surface state of TI  via
proximity effect has been verified by the presence of supercurrent
through the Josephson junctions on TI \cite{Sacepe11,DMZhang,Veldhorst,Williams12,Snelder}. Josephson current in superconductor(S)/ferromagnetic insulator(FI)/S junction
on TI stimulates us
since anomalous CPR $I(\varphi) \sim \sin(\varphi -\varphi_{0})$
discussed in conventional S/Ferromagnet (F)/S junction without  TI
\cite{Grein,Eschrig07,Nazarov07,Asano07,Eschrig08,BuzdinPRL,KonschellePRL}
can be realized easily.
It is noted that $\varphi_{0}$
can be tunable by magnetization \cite{Tanaka2009}.

We can imagine that more dramatic features will be expected in
$d$-wave superconductor(DS) /FI/DS Josephson junctions on TI.
Although there have been several works about DS/FI/DS junctions
\cite{Linderprl10,Linderprb10,Tafuri}, CPR has not been clarified
for general orientations of the junctions.
Recent experiments have shown that
induced gap function on the surface of TI which is formed on
high-$T_{c}$ cuprate is almost isotropic.
This shows that the induced pair potential  has
predominant $s$-wave symmetry\cite{fullygaped}. The possibility of inducing
$d$-wave pairing on TI surface in the actual experiment
is still on debate now \cite{Yao2014,SYXU,Lee}.
Therefore,
besides the CPR in $d$-wave superconductor junctions, we must study
the CPR in junctions with $s$+$d$-wave symmetry
for comparison with actual experiments.

In this paper, in order to calculate DC Josephson current,
we develop a formalism of Green's function of quasiparticles on the surface of TI
\cite{McMillan}.
Both Josephson current and local density of states (LDOS) can be calculated for general orientations
of junctions. In general,
the obtained current phase relation $I(\varphi)$
has a complex $\varphi$ dependence.
$I(\varphi)=-I(-\varphi)$ is easily to be broken by magnetization in FI
and $I(\varphi)$ can not be simply expressed  by
$I_{0}\sin(\varphi-\varphi_{0})$ with nonzero
$\varphi _{0}$.
The extreme case is $d_{x^{2}-y^{2}}$/FI/$d_{xy}$-wave junctions, where
CPR becomes $\sin 2\varphi$ without magnetization
due to the absence of
the first order Josephson coupling.
If we switch on magnetization, exotic CPR becomes possible depending
on the direction of magnetization in FI:
i)mixture of  $\cos \varphi$ and $\sin(2\varphi)$ terms
with $I(\varphi)=-I\left( -\varphi +\pi
\right)$ and
ii)$\sin (2\varphi-2\varphi_{0})$.
The complex CPR $I(\varphi)=-I\left( -\varphi +\pi
\right)$ with $2\pi$ periodicity in i) is not realized in the preexisting
high-$T_{c}$ cuprate junctions without TI \cite{Tsuei,Tanaka2000,GKI}.
We also calculate Josephson current where $s$-wave and $d$-wave pair potentials mix.
It is found that the anomalous CPR with
$I\left( \varphi =0\right) \neq 0$ exists for
junctions of asymmetric orientations even if
$s$-wave component becomes dominant.
This feature serves as a guide to detect the
proximity induced $d$-wave component of pair potential
on the surface of TI.

%In the experimental pursuit of d-wave pair potential on the TI surface by growi%ng Bi-family TI on top of
%high-Tc cuprate crystal, it shows that the induced gap function has
%predominant s-wave symmetry\cite{fullygaped}. The possibility of inducing
%d-wave pairing on TI surface is still on debate\cite{SYXU,Lee}. Therefore,
%besides the d-wave superconductor, we will also consider s+d wave pairing in
%the same manner. It is found that the anomalous CPR $\left\vert
%I\left( \varphi \right) \right\vert \neq \left\vert I\left( -\varphi \right)
%\right\vert $ can also be obtained. And significantly, this phenomena shows $%
%I\left( \varphi =0\right) \neq 0$ which can be applied to probe the fingerprint% of the induced d-wave pair potential
%in the high-Tc cuperate/TI hybrid systems.

The outline of this paper is as follows.
In section II, we present the
model and derive the retarded Green's function and Josephson current. In
section III, we show numerical results of Josephson current in DS/FI/DS
junctions. In section IV, we show the corresponding results of Josephson junctions when
$s$-wave component is induced in DS.
A concluding remarks will be given in section V.

\section{II, Model and formulas}

\begin{figure}[ptbh]
\begin{center}
\includegraphics[width = 60 mm]{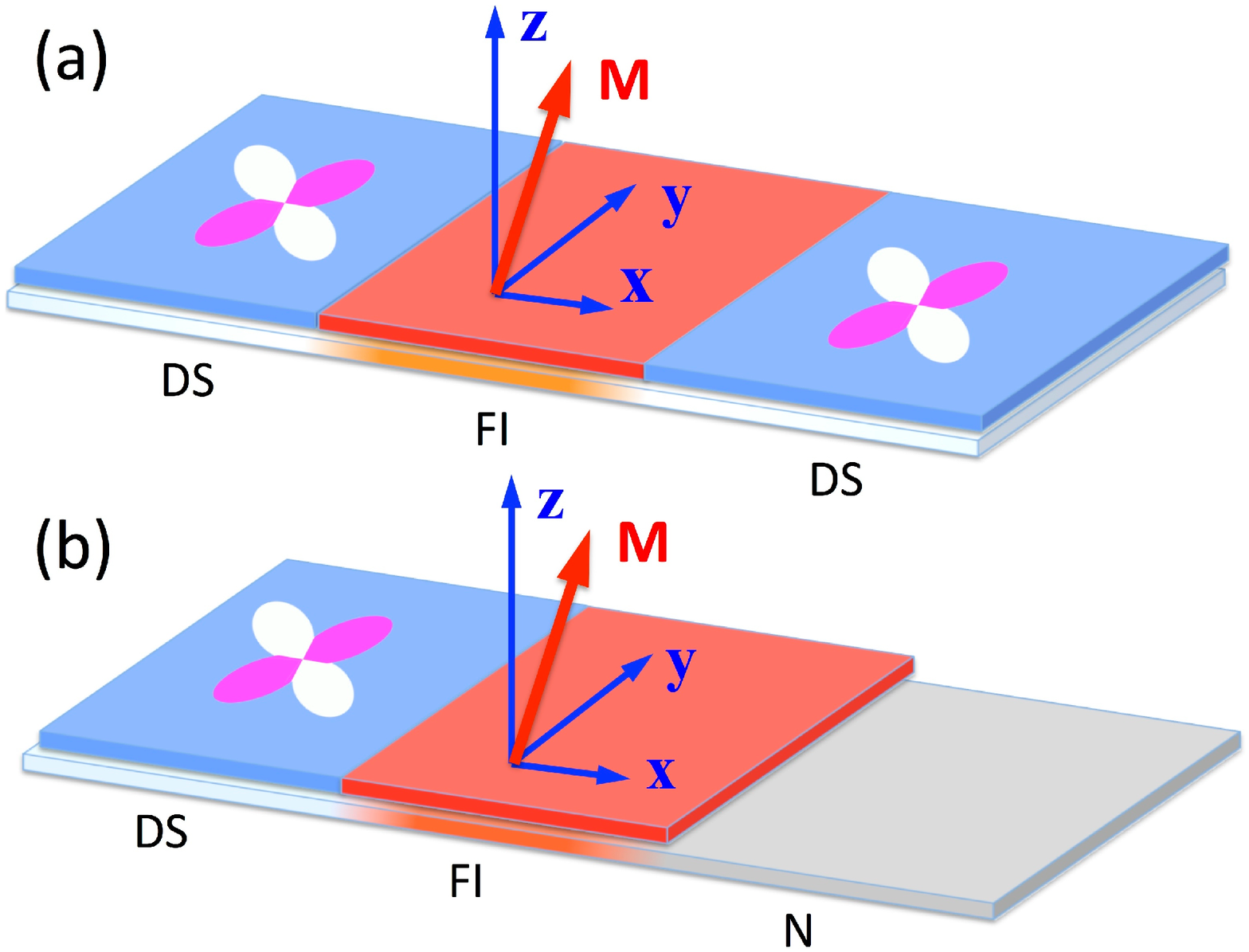}
\end{center}
\caption{Schematics of the system: (a) $d$-wave superconductor (DS)/ferromagnetic
insulator (FI)/DS Josephson junction and (b) DS/FI/normal metal (N) junction on the surface
of a 3D topological insulator.}
\label{fig1}%
\end{figure}

As depicted in Fig.\ref{fig1}(a), we consider a DS/FI/DS junction on a 3D TI surface.
The effective Hamiltonian for the BdG equations is given by
\begin{equation}
\mathcal{H}=\left[
\begin{array}{cc}
h(k_{x},k_{y})+M & i\hat{\sigma}_{y}\Delta \left( \theta \right) \\
-i\hat{\sigma}_{y}\Delta ^{\ast }\left( \theta \right) & -h^{\ast
}(-k_{x},-k_{y})-M^{\ast }%
\end{array}%
\right] ,  \label{Hamiltonian}
\end{equation}%
where $h(k_{x},k_{y})=\hbar v_{f}(k_{y}\hat{\sigma}_{x}-k_{x}\hat{\sigma}%
_{y})-\mu (\Theta \left( -x\right) +\Theta \left( x-L\right) )$. $\hat{\sigma%
}_{x,y,z}$ is the Pauli matrix in the spin space and $\mu $ is the
chemical potential in the superconducting region. The exchange field in FI region is $M=\sum_{i=x,y,z}m_{i}%
\hat{\sigma}_{i}\Theta \left( x\right) \Theta \left( L-x\right) $\cite%
{Tanaka2009}. The pair potential is given by $\Delta _{0}\left( T\right)
\cos \left( 2\theta -2\chi _{1}\right) \Theta (-x)+\Delta _{0}\left(
T\right) \cos \left( 2\theta -2\chi _{2}\right) e^{-i\varphi }\Theta (x-L)$,
where $\varphi $ and $\theta $ are the macroscopic superconducting phase and
the propagating angle, respectively. The quantity $\chi $ is taken to be the
angle between the $x$-axis and the $a$-axis of  $d$-wave superconductor
 on top of
the TI surface. $\Delta _{0}\left( T\right) $ is assumed to obey the BCS
relation $\Delta _{0}(T)=\Delta _{0}\tanh (1.74\sqrt{T_{c}/T-1})$ with $%
\Delta _{0}=1.76k_{B}T_{c}$ and $T_{c}$ is the critical temperature.

To construct the retarded Green's function, we first seek the solutions for
the four types of quasiparticle injection processes: left injection for
electron (hole): $\psi _{1}\left( \psi _{2}\right) $ and right injection for
electron (hole): $\psi _{3}\left( \psi _{4}\right) $. Because of the translational invariance along $y$-axis, the wave functions  $\psi
_{1(2)}=\psi _{1(2)}\left( x\right) e^{ik_{y}y}$ in the left superconducting
region can be expressed as
\begin{subequations}
\begin{alignat}{4}
& \psi _{1}(x)=\hat{A}_{1}e^{ik_{x}x}+a_{1}\hat{A}_{4}e^{ik_{x}x}+b_{1}\hat{A%
}_{3}e^{-ik_{x}x}, & & & & & & \\
& \psi _{2}(x)=\hat{A}_{2}e^{-ik_{x}x}+a_{2}\hat{A}_{3}e^{-ik_{x}x}+b_{2}%
\hat{A}_{4}e^{ik_{x}x}, & & & & & &
\end{alignat}%
where $k_{x}=\mu \cos \theta /\hbar v_{F}$. Here the magnitudes of the
momenta for electrons and holes are approximated to be equal since we
have made the assumption of $E$, $\Delta \ll \mu $. The spinors are given by
$\hat{A}_{1}=(i,e^{i\theta },-e^{i\theta }\gamma _{1},i\gamma _{1})^{T}$,
$\hat{A}_{2}=(ie^{i\theta }\gamma _{2},-\gamma _{2},1,ie^{i\theta })^{T}$,
$\hat{A}_{3}=\left( ie^{i\theta },-1,\gamma _{2},ie^{i\theta }\gamma
_{2}\right) ^{T}$ and $\hat{A}_{4}=(i\gamma _{1},e^{i\theta }\gamma
_{1},-e^{i\theta },i)^{T}$ with $\gamma _{1\left( 2\right) }=\Delta
_{1\left( 2\right) }/(E+\sqrt{E^{2}-\Delta _{1\left( 2\right) }^{2}})$ and $%
\Delta _{1\left( 2\right) }=\Delta _{0}\cos \left( 2\theta \mp 2\chi
_{1}\right) $. Other wave functions can be solved in a similar way. The
retarded green's function $G^{r}(x,x^{\prime },y,y^{\prime
})=\sum\nolimits_{k_{y}}G_{k_{y}}^{r}\left( x,x^{\prime }\right)
e^{ik_{y}(y-y^{\prime })}$ can be obtained by combing all the injection
processes\cite{McMillan}:
\end{subequations}
\begin{equation}
G_{k_{y}}^{r}\left( x,x^{\prime }\right) =\left\{
\begin{array}{c}
\alpha _{1}\psi _{1}\left( x\right) \tilde{\psi}_{3}^{T}\left( x^{\prime
}\right) +\alpha _{2}\psi _{1}\left( x\right) \tilde{\psi}_{4}^{T}\left(
x^{\prime }\right) + \\
\alpha _{3}\psi _{2}\left( x\right) \tilde{\psi}_{3}^{T}\left( x^{\prime
}\right) +\alpha _{4}\psi _{2}\left( x\right) \tilde{\psi}_{4}^{T}\left(
x^{\prime }\right) , \\
(x>x^{\prime }), \\
\beta _{1}\psi _{3}\left( x\right) \tilde{\psi}_{1}^{T}\left( x^{\prime
}\right) +\beta _{2}\psi _{4}\left( x\right) \tilde{\psi}_{1}^{T}\left(
x^{\prime }\right) + \\
\beta _{3}\psi _{3}\left( x\right) \tilde{\psi}_{2}^{T}\left( x^{\prime
}\right) +\beta _{4}\psi _{4}\left( x\right) \tilde{\psi}_{2}^{T}\left(
x^{\prime }\right) , \\
(x<x^{\prime }),%
\end{array}%
\right.
\end{equation}%
where $\tilde{\psi}_{i=1\sim 4}$ are the corresponding conjugated processes
of $\psi _{i=1\sim 4}$. The coefficients $\alpha _{i=1\sim 4}$ and $\beta
_{i=1\sim 4}$ are determined by satisfying the boundary conditions for all $x
$, $x^{\prime }$ across the regions:%
\begin{equation}
G_{k_{y}}^{r}(x+0,x)-G_{k_{y}}^{r}(x-0,x)=\hbar ^{-1}v_{f}^{-1}(i\hat{\tau}%
_{z}\hat{\sigma}_{y}),
\end{equation}%
where $\hat{\tau}_{x,y,z}$ is the Pauli matrix in the particle-hole space.
The dc Josephson current for DS/FI/DS junction is determined by electric
charge conservation rule
\begin{equation}
\partial _{t}P+\partial _{x}J_{x}+S=0,
\end{equation}%
where $P=e(\Psi _{\uparrow }^{\dag }\Psi _{\uparrow }+\Psi _{\downarrow
}^{\dag }\Psi _{\downarrow })$, $J_{x}=iev_{f}(\Psi _{\uparrow }^{\dag }\Psi
_{\downarrow }-\Psi _{\downarrow }^{\dag }\Psi _{\uparrow })$ and $S=2e%
\mathrm{Im}[\Delta ^{\ast }\Psi _{\downarrow }\Psi _{\uparrow }-\Delta
^{\ast }\Psi _{\uparrow }\Psi _{\downarrow }]$ are electric charge density,
electric current and source term, respectively. After straightforward
derivation following Ref.\cite{FT}, we find that the total Josephson
current is given by%
\begin{equation}
I_{x}=\frac{ek_{B}T}{2\hbar }\sum\limits_{k_{y},\omega _{n}}\mathrm{sgn}%
(\omega _{n})[\frac{\Delta _{1}a_{1}(i\omega _{n})}{\sqrt{\omega
_{n}^{2}+\Delta _{1}^{2}}}-\frac{\Delta _{2}a_{2}(i\omega _{n})}{\sqrt{%
\omega _{n}^{2}+\Delta _{2}^{2}}}].  \label{ft}
\end{equation}%
$a_{1\left( 2\right) }(i\omega _{n})$ is obtained by analytical continuation
$E$ to $i\omega _{n}$, where $\omega _{n}$ is the Matsubara frequency $%
\omega _{n}=\pi k_{B}T(2n+1),(n=0,\pm 1,\pm 2....)$. Eq.(\ref{ft}) looks similar to the
extended Furusaki-Tsukada's formula\cite{FT} for anisotropic $d$-wave pair
potential\cite%
{TanakaD96,TanakaD97}. In addition, Eq.(\ref{ft}) is also applicable to the
Josephson current of $s$+$d$ wave pairing in which one substitutes $\Delta
_{1\left( 2\right) }$ by $\Delta _{0}+\eta \Delta _{0}\cos \left( 2\theta
\mp 2\chi _{1}\right) $ of which $\eta \geq 0$ is the ratio between $d$-wave
pairing and $s$-wave pairing.

\section{III, Josephson Effect in DS/FI/DS Junction}

\begin{figure}[ptbh]
\begin{center}
\includegraphics[width = 86 mm]{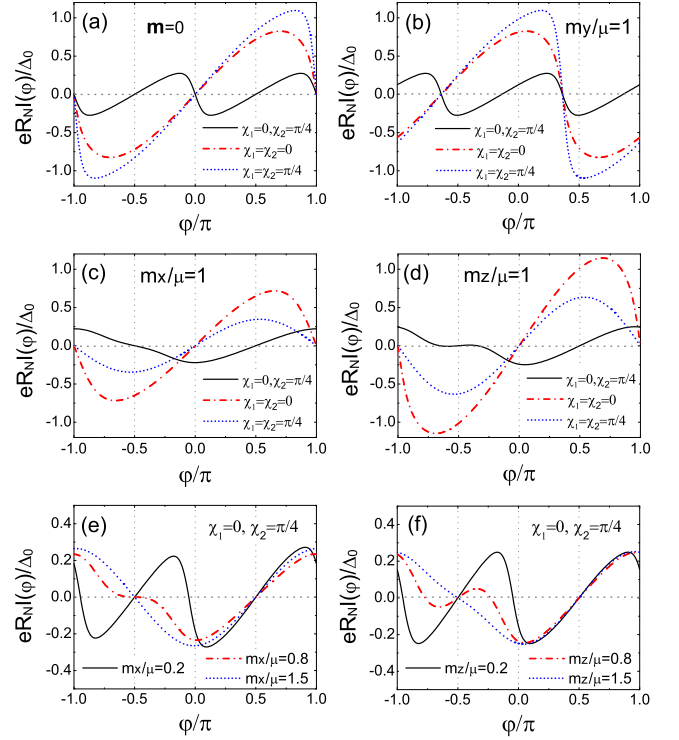}
\end{center}
\caption{Josephson currents as a function of $\varphi $ in the DS/FI/DS junctions.
Magnetization in FI is (a) zero, (b) along $x$%
-axis, (c) along $y$-axis and (d) along $z$-axis.
Three geometries are considered in panels (a)$\sim$(d): $d_{x^{2}-y^{2}}$/FI/$%
d_{xy}$, $d_{x^{2}-y^{2}}$/FI/$d_{x^{2}-y^{2}}$ and $d_{xy}$/FI/$d_{xy}$.
(e) Josephson currents in $d_{x^{2}-y^{2}}$/FI/$d_{xy}$ junctions with
$m_{x}/\mu =0.2$, $0.8,$ and $1.5$
and (f) those with $m_{z}/\mu =0.2$, $0.8$, and $1.5$.
Other parameters are set as $T=0.05T_{c}$, $\mu =1$, $\hbar v_{f}=1$, $%
\Delta =0.01$ and $L=1$.
}
\label{fig2}%
\end{figure}

In this section, we show the results of Josephson current $I$ in DS/FI/DS
junctions, which has been normalized to $eR_{N}I/\Delta _{0}$ where $R_{N}$
is the interface resistance per unit area in the normal state. To analyze
the CPR further, we decompose the Josephson current into a series of
different orders of Josephson coupling,
\begin{equation}
I\left( \varphi \right) =\sum_{n}
I_{n}\sin \left( n\varphi \right) +J_{n}\cos \left(
n\varphi \right),
\end{equation}
where $n\geq 1$ is an integer. Fig.\ref{fig2}(a) shows CPR without magnetization. In this case, the CPR is expressed as
$\sum_{n} I_{n}\sin \left( n\varphi \right) $ and $J_{n}$ is zero.
In the condition with $\chi _{1}=0$ and $\chi _{2}=\pi /4$,
the CPR $I\left( \varphi \right)$
becomes $\sum_{n} I_{n}\sin \left( n\varphi \right) $ $\left(
n=2,4...\right) $.
The feature of this CPR is the same as that in the standard
$d$-wave junctions without TI with the pair potential considered here.
However, as the magnetization
switches on, the CPR dramatically  changes.
Figure \ref{fig2}(b) shows
that $m_{y}$ gives a shift of  phase difference,
which is similar to $\varphi _{0}$-junctions
realized in conventional $s$-wave superconductor/ferromagnet hybrid systems \cite{Grein,Eschrig07,Nazarov07,Asano07,Eschrig08,BuzdinPRL,KonschellePRL}.
As the magnetization along $m_{x}$ or $m_{z}$-axis
appears, the
qualitative features of CPR of symmetric $d%
_{x^{2}-y^{2}}$/FI/$d_{x^{2}-y^{2}}$ and $d_{xy}$/FI/$d_{xy}$ junctions
do not change as compared to the case without magnetization as
shown in dotted and dash-dotted lines of Figs. \ref{fig2}(c) and (d).
However, in the asymmetric
$d_{x^{2}-y^{2}}$/FI/$d_{xy}$ junction, the CPR is quite anomalous and the
component proportional to
$\sum_{n}J_{n}\cos \left( n\varphi \right) $ is generated. We find that
$I\left( \varphi \right) $ can be expressed by
 $\sum_{k} [I_{2k}\sin \left( 2k\varphi
\right) +J_{2k-1}\cos (2k-1)\varphi]$ where $k\geq 1$ is an integer, and therefore, $I(\varphi)$ becomes zero at
$\varphi=\pm \pi/2$ as shown in solid lines in Figs. \ref{fig2}(c) and (d). The present CPR is completely different from that of the standard
$d_{x^{2}-y^{2}}$/FI/$d_{xy}$ junction without TI.
We can see that the term proportional to $J_{1}\cos \left( \varphi \right)
$ becomes dominant in the limit of large $m_{x}$ or
$m_{z}$ in
Figs.\ref{fig2}(e) and (f).

To explain the anomalous CPR for
nonzero $m_{x}$ or $m_{z}$ in $d_{x^{2}-y^{2}}$/FI/$d_{xy}$
junction on TI surface, we focus on the symmetry of this Hamiltonian.
%%%%
We consider the mirror reflection symmetry with respect to $xz$-plane, $%
M_{xz}=i\sigma _{y}\tau _{0}$, and the time-reversal symmetry, $T=-i\sigma
_{y}\mathcal{K}\tau _{0}$, where $\mathcal{K}$ is the complex conjugation
operator. In the present system, both symmetries are broken. However, since the pair potential of $d_{x^{2}-y^{2}}$ ($d_{xy}$) is mirror even
(odd) with respect to $xz$-plane, $M_{xz}$ operation produces additional phase, $I\left(
\varphi \right) \rightarrow I\left( \varphi +\pi \right) $. It is also known
that time reversal operation transforms $I\left( \varphi \right) $ to $-I\left(
-\varphi \right) $. Hence, the composition operator $\tilde{T}=M_{xz}T$ will
give rise to $I\left( \varphi \right) \rightarrow -I\left( -\varphi +\pi
\right) $. Taking into account the fact that $\tilde{T}$ makes the $\left(
k_{x},k_{y}\right) $ state to the $\left( -k_{x},k_{y}\right) $ one, we can
arrive at
\begin{equation}
\tilde{T}\mathcal{H}\left( -i\partial _{x},k_{y},\varphi \right) \tilde{T}%
^{-1}\rightarrow \mathcal{H}\left( i\partial _{x},k_{y},-\varphi +\pi
\right) .  \label{symm}
\end{equation}%
It means that $-I\left( -\varphi +\pi \right) =I\left( \varphi \right) $
will be satisfied at any $\varphi$ if we consider the junctions between a mirror even and mirror odd pair potential.
In the $d_{x^{2}-y^{2}}$/FI/$d_{xy}$ junction with $m_{x}$ or $m_{z}$, we
can find that relation (\ref{symm}) fulfills at any $\varphi$, which
indicates $I\left( \varphi =\pm \pi /2\right) =0$.
Above analysis based on mirror reflection symmetry
has been applied in the Josephson junctions between a singlet and triplet superconductor\cite{Yip2009}.
Now, let us look at the Josephson current at $\varphi=0$. In the standard DS/FI/DS junctions without TI substrate, due to the
spin SU(2) symmetry, the rotation or mirror reflection of the ferromagnetism
does not change the CPR and one can always find $I(\varphi=0)=0$\cite{note1}.
However, this SU(2) symmetry is broken on TI surface due to its nature of spin-momentum locking and thus $I(\varphi=0)$ becomes nonzero
which generates exotic $2\pi$-periodic CPR $-I(-\varphi+\pi)=I(\varphi)$.
\begin{figure}[ptbh]
\begin{center}
\includegraphics[width = 82 mm]{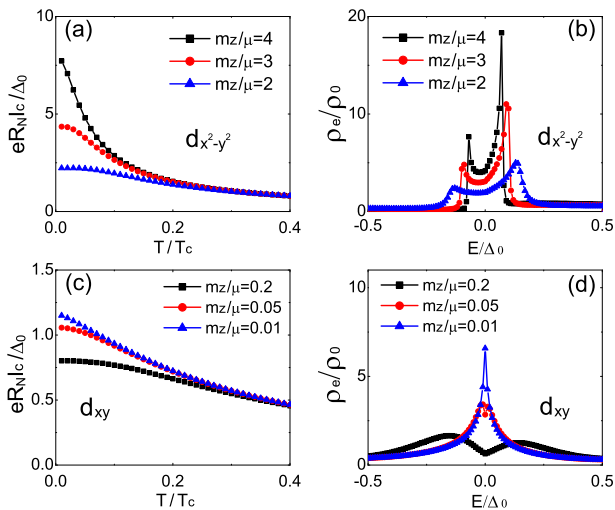}
\end{center}
\caption{(a) The maximum Josephson currents in the $d_{x^{2}-y^{2}}$/FI/$d_{x^{2}-y^{2}}$
junctions. (b) The LDOS on the surface of DS in the $d_{x^{2}-y^{2}}$/FI/N
junctions. $\rho_{0}$ is the electron density of states of the bulk N at Fermi energy. (c) and (d) are the maximum Josephson currents and LDOS in $d_{xy}$/FI/$d_{xy}
$ junctions, respectively. The magnetization is along $z$-axis in all panels. Other parameters are set as the same as Fig.\ref{fig2}.
}
\label{fig3}%
\end{figure}

Next, we plot the temperature dependence of the maximum Josephson current $I_{c}$ of
DS/FI/DS junctions in the left panels of Fig.\ref{fig3}. For simplicity, only the $z$ component of magnetization $m_{z}$ is considered.
We concentrate on the low temperature
region $T/T_{c}\leq 0.4$ in which the behavior of $I_{c}$ is highly
influenced by the zero-energy states (ZESs). Therefore, we display the
surface density of states at the edge of DS. It is obtained by calculating
the LDOS $
\rho _{e}(x,E)=-\frac{1}{\pi }\sum_{k_{y}}\mathrm{Im}%
[G_{k_{y},11}^{r}(x,x,E)+G_{k_{y},22}^{r}(x,x,E)]$ at the DS/FI interface in the DS/FI/N junction as illustrated in
Fig.\ref{fig1}(b). From Fig.\ref{fig3}(a), we can see that temperature dependence of $I_{c}$ in $d_{x^{2}-y^{2}}$/FI/$d%
_{x^{2}-y^{2}}$ junctions changes from Kulik-Omelyanchuk (K-O)\cite{KO}
type to Ambegaokar-Baratoff (A-B)\cite{AB} type with decreasing $m_{z}$.
However, in  $d_{xy}$/FI/$d_{xy}$ junction, this tendency is reversed
when we decrease $m_{z}$, as shown in Fig.\ref{fig3}(c).
As seen from  Figs.\ref{fig3}(b) and (d), we can see that as LDOS at zero energy is enhanced, temperature dependence of $I_{c}$ is reduced to be
the K-O type. In other case, it is in the A-B type.
Since the spin degeneracy is
lifted on the surface states of TI  by spin momentum locking,
we obtain the highly asymmetric
Yu-Shiba-Rusinov type of LDOS \cite{Yu,Shiba,Rusinov} in the $d_{x^{2}-y^{2}}$/FI
interface. This finding is similar to that in the $s$-wave superconductor/FI/N junction\cite{LuSUST}.

\section{IV, Josephson effect with $s$+$d$-wave pairing}
\begin{figure}[ptbh]
\begin{center}
\includegraphics[width = 75 mm]{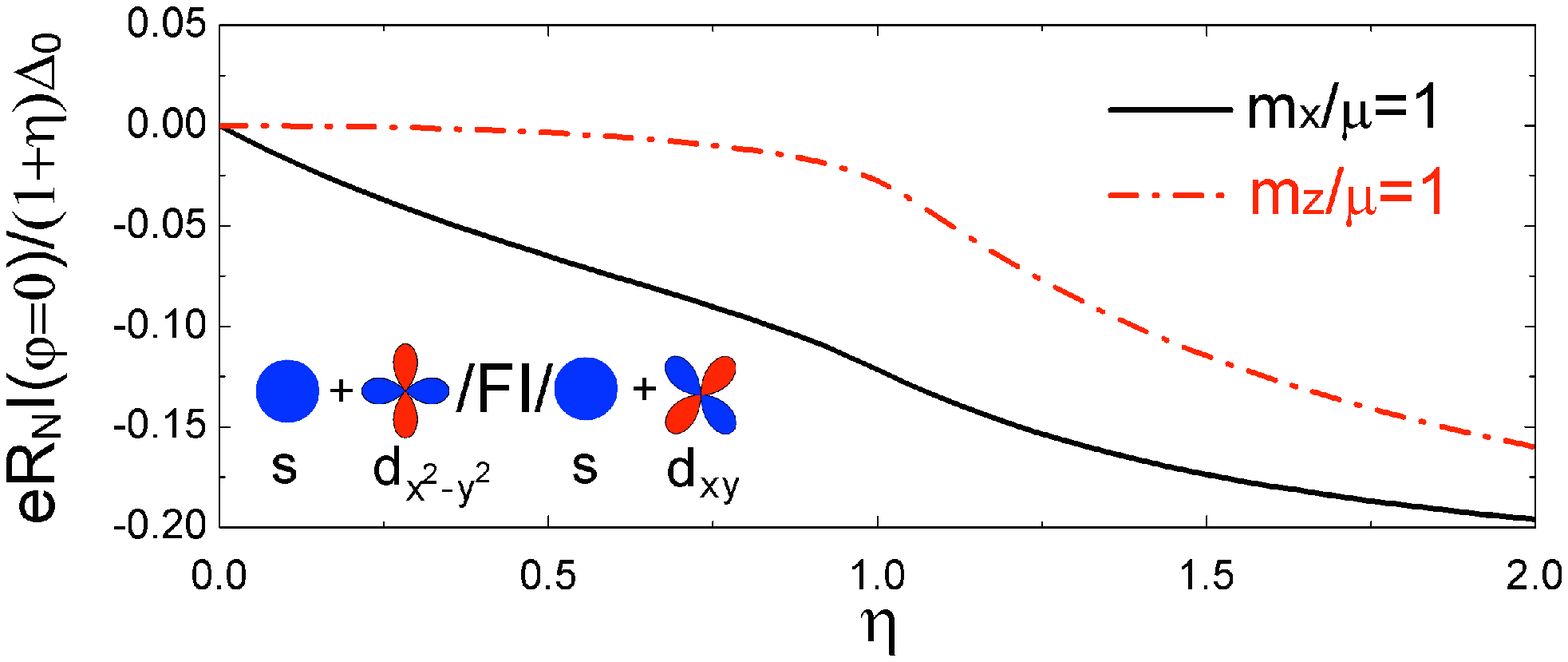}
\end{center}
\caption{$I(\varphi=0)$ as a function of $\eta $ in $s$+$d_{x^2-y^2}$/FI/$s$+$d_{xy}$ junctions on TI. Other parameters are set as the same as in Fig.\ref{fig2}.
}
\label{fig6}%
\end{figure}

Recent experiments have shown that
the induced energy gap by high $T_{c}$ cuprate on the surface of TI is almost isotropic\cite{fullygaped}.
It is interesting to clarify the
role of the induced $s$-wave pair potential on DS/FI/DS junctions.
In this section, we calculate Josephson current in
$s$+$d$-wave/FI/$s$+$d$-wave junctions on TI.
The pair potential is $\Delta=\Delta _{0}+\eta \Delta _{0}\cos( 2\theta
- 2\chi) $ with $\chi=0(\pi/4)$ on the left(right) side. The ratio $\eta$ is chosen to be
$0.5$ so that the system is $s$-wave
dominant and fully gapped.
The obtained
$I(\varphi)$ has a typical sinusoidal shape of
 $s$-wave Josephson current where the first order
coupling $I_{1}\sin \varphi $ plays the predominant role.
Because of the $s$-wave component of pair potential,
the mirror reflection symmetry
$\tilde{T}$ at $\varphi =\pm \pi /2$ is broken and thus
nonzero current $I(\varphi=\pm \pi /2)$ can be expected.
Also, in the presence of $m_{x}$ or $m_{z}$, we
find a nonzero Josephson current at $\varphi =0$ $,\pi$ in such junctions.
%For  $s$+$d_{xy}$/FI/$s$+$d_{xy}$
%case with $\varphi=0$, the system is symmetric with respect to the $yz$-plane at the center of FI when the
%magnetization axis is along $z$-axis. Thus $I(\varphi=0,\pi)$ becomes zero.
%If the magnetization is along $x$-axis, $I(\varphi =0$ $,\pi )\neq 0$ due to
%the absence of this symmetry.
The obtained anomalous CPR in the $s$+$d$-wave Josephson
junctions can be used to probe the $d$-wave component of the
induced pair potential on TI surface. For example, one can
observe the supercurrent flow without macroscopic phase
difference in  $s$+$d_{x^{2}-y^{2}}$%
/FI/$s$+$d_{xy}$ junctions.
As seen in Fig.\ref{fig6}, the existence of $d$-wave component
generates a nonzero current $I(\varphi=0)$ when one
turns on either $m_{x}$ or $m_{z}$.

\section{V, Conclusion}
In summary, we have theoretically studied the Josephson effect in $d$-wave
superconductor-ferromagnet insulator (FI) hybrids on the surface of TI.
Depending on the orientation of the magnetization in FI, the exotic
current-phase relation which violates $I(\varphi )\neq -I(-\varphi )$ have
been obtained in two different ways: (i) through a simple phase shift and
(ii) mixture of $\cos \varphi $ term into the original CPR.
The latter case can generate the exotic current-phase relation $I(\varphi)=-I(-\varphi+\pi)$ with $2\pi$ periodicity.
We show that the Josephson
current is enhanced due to the zero energy states on the edge of $d$-wave
superconductor. For comparison with actual experiments, we calculate the
Josephson current when both $s$- and $d$-wave pair potentials exist. The
anomalous current-phase relation is also found which provides a way to probe
the fingerprint of $d$-wave pair potential in high-$T_{c}$ cuprate/TI
heterostructures. Our preliminary theoretical investigation has practical
significance for controlling the Josephson current and designing new
functional devices.
\section{ACKNOWLEDGEMENTS} We thank Y. Asano, M. Sato and P. Burset for
valuable discussions. This work
was supported in part by Grants-in-Aid for Scientific
Research from the Ministry of Education, Culture,
Sports, Science and Technology of Japan (Topological
Quantum Material No.15H05853) and by the
Ministry of Education and Science of the Russian Federation Grant No.14Y.26.31.0007.


\begin{thebibliography}{9}
\bibitem{note1} In cuprate junctions where the
geometry is the same as Fig.1(a), the orientation of ferromagnetism plays
no role in CPR due to the spin SU(2) symmetry. Therefore in the case $\varphi =0$%
, the system and its time reversal copy have the same Josephson current,
which implies $I\left( \varphi =0\right) =0$.
\end{thebibliography}
\end{document}